\documentclass[a4paper,11pt]{article}
\usepackage{pos}
\usepackage{hyperref}
\usepackage{subcaption}
\usepackage{wrapfig}

\title{Bayesian inference of 3D densities of galactic HI and H2}
 \ShortTitle{Bayesian inference of 3D densities of galactic HI and H2}

\author*[a]{Laurin S\"oding}
\author[a]{Philipp Mertsch}
\author[a]{Vo Hong Minh Phan}
\affiliation[a]{Institute for Theoretical Particle Physics and Cosmology (TTK), RWTH Aachen University, 52056 Aachen, Germany}

\emailAdd{soeding@physik.rwth-aachen.de}
\emailAdd{pmertsch@physik.rwth-aachen.de}
\emailAdd{vhmphan@physik.rwth-aachen.de}

\abstract{Due to our vantage point in the disk of the Galaxy, its 3D structure is not directly accessible. However, knowing the spatial distribution, e.g. of atomic and molecular hydrogen gas is of great importance for interpreting and modelling cosmic ray data and diffuse emission. Using novel Bayesian inference techniques, we reconstruct the 3D densities of atomic and molecular hydrogen in the Galaxy together with (part of) the galactic velocity field. In order to regularise the infinite number of degrees of freedom and obtain information in regions with missing or insufficient data, we incorporate the correlation structure of the gas fields into our prior. Basis for these reconstructions are the data-sets from the HI4PI-survey on the 21-cm emission line and the CO-survey compilation by Dame et al. (2001) on the ($1\rightarrow0$) rotational transition together with a variable gas flow model. We present the preliminary estimated mean surface mass densities and corrections to the prior assumption of the galactic velocity field. In the future, we plan to relax assumptions on the optical thickness and include additional data to further constrain either the galactic velocity field or the gas densities.}

\ConferenceLogo{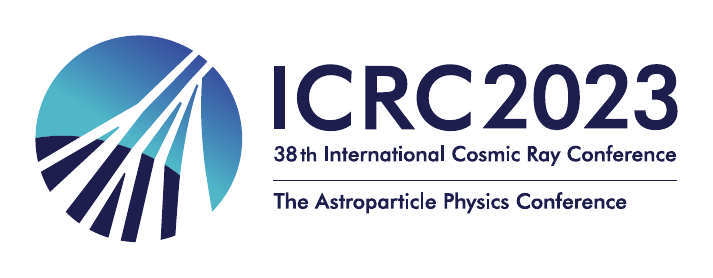}

\FullConference{%
38th International Cosmic Ray Conference (ICRC2023)\\
  26 July - 3 August, 2023\\
  Nagoya, Japan}

\begin{document}
\maketitle

\section{Introduction}
In order to properly interpret measurements of cosmic rays and gamma-ray diffuse emission, it is necessary to understand the emission, propagation and absorption of radiation in the interstellar medium of the Milky Way. This medium is a  complex system consisting mainly of gas, magnetic fields, interstellar radiation fields and cosmic rays which permeate the entire Galaxy. While it makes up for only a few percent of the total mass of the Galaxy (the majority is in the form of stars or dark matter), it fills out most of the available volume and thereby defines the dynamics of radiation and particles within.

Due to our vantage point, the 3D-distribution of the constituents of the Galaxy is not easily determined by observations of the sky. No matter in which direction we point our telescopes, we always observe an integrated signal of radiation that has travelled an a priori unknown distance through the Galaxy.
However, due to galactic rotation and peculiar motion, light that reaches us will be Doppler-shifted by a certain amount, depending on the relative velocity of its source with respect to us. This is particularly useful when looking at emission lines that have a narrow width as it enables us to determine the relative velocity of its emitter and observer very precisely. This idea is unfortunately somewhat tainted by the fact that we do not know the precise structure of the galactic velocity field - and even circular rotation features a velocity ambiguity for positions within the solar circle.
Any attempt to produce 3D maps of some quantity from such data will thus have to specify some rule according to which said quantity is placed when there is an ambiguity. Multiple approaches have been tried, most of them treating every line-of-sight (direction) independently, thereby missing out on a lot of information.
This work will attempt to produce 3D maps of the distribution of HI (atomic hydrogen) and H$_2$ (molecular hydrogen) in the Milky Way using novel Bayesian inference techniques, exploiting spatial correlations of the gas structure to regularise ambiguities. Our approach will not only yield maps of the estimated gas densities, but also uncertainty information.

The two observational datasets used are that of the HI4PI-survey \cite{HI4PI} mapping the 21-cm emission of atomic hydrogen in the galaxy (see figure \ref{fig:HIdata}) and the CO-survey compilation by \cite{DameCO} observing the $1 \rightarrow 0$ rotational transition of CO as a tracer for molecular clouds and thereby H$_2$ (see figure \ref{fig:COdata}).

This work builds on precursory reconstructions (see \cite{MertschCO, MertschHI}) with some key differences:
\begin{enumerate}
    \item A different numerical grid is used trading resolution far away from the observer for a much more refined resolution nearby.
    \item The inference of galactic HI and H$_2$ is unified into a common inference process coupled by a common galactic velocity field.
    \item The galactic velocity field is partly inferred, modifying our prior assumption by adding a curl-free field.
\end{enumerate}
In the following section, we will formulate this problem in a Bayesian manner and shortly describe the used approach to this very-high-dimensional problem. Thereafter, we will show our preliminary results, i.e. 3D-maps of the distribution of HI and H$_2$ in the galaxy.
\section{Method}
\subsection{Bayesian formulation}
In the language of probabilities, we want to know the probability of the gas distribution in the galaxy (called signal $s$) given the data $d$ obtained by the sky surveys. Using Bayes' law, this can be written as
\begin{equation}
    P(s|d) = \frac{P(s,d)}{P(d)} = \frac{P(d|s)P(s)}{P(d)}\,.
    \label{eq:Bayes}
\end{equation}
Since the datasets we are using measured the brightness temperature $T$ (a measure of the intensity of the observed radiation) as a function of relative line-of-sight velocity and position on the sky, we will attempt to infer the CO and HI volume emissivity $s = (\varepsilon_\text{HI}(\Vec{x}), \varepsilon_\text{CO}(\Vec{x}))$ simultaneously and later convert to gas densities.

Equation \ref{eq:Bayes} is often solved by creating a model that allows to sample from the prior distribution and compute the likelihood of said sample. Then, algorithms like MCMC sampling can be used to probe the shape of the posterior probability. For problems with many free parameters (usually more than ${\gtrsim}10^2$), this becomes computationally unfeasible.
A commonly used approach for high-dimensional problems are so-called \textit{Variational Inference}(VI)-methods \cite{VIReview}. The idea of these methods is to approximate the posterior by a family of parametric distributions, for example a multi-variate Gaussian distribution.

The parameters of this approximation can be determined by minimising the ''distance'' between the approximated posterior and the true posterior, for example via the Kullback-Leibler-divergence \cite{KullbackLeibler}. Computing this in theory involves the inversion of the full covariance matrix of all the correlated latent variables which - with millions or more of free parameters - is impossible even to store in common memory modules. As an approximation, it has been suggested to replace the inverse of the full covariance matrix by the inverse Fisher information metric, an approach known as Metric Gaussian Variational Inference (MGVI, \cite{MGVI}). This method can be applied to problems with more than $10^6$ parameters while still being computationally efficient on regularly available hardware. 

This algorithm is implemented in an iterative scheme in the publicly available code-package \textsc{nifty8}\footnote{Available at \url{https://gitlab.mpcdf.mpg.de/ift/NIFTy}}. It alternates between estimating the covariance of the probability distribution with the inverse Fisher information metric at the current mean and optimising the mean of the distribution by minimising the Kullback-Leibler divergence to the true posterior with respect to the mean. 
This does not require explicitly computing the covariance matrix at any point (which would require the inversion of the Fisher information metric): instead the Kullback-Leibler divergence is estimated stochastically by drawing samples from a Gaussian with the appropriate covariance, leading to linear scaling in the model parameters. This can be implemented in terms of implicit operators which apply the Fisher information metric to some vector and then solving a linear system via conjugate gradient methods to obtain the application of the inverse Fisher information metric to some vector (which then features the desired correlation structure).

This way, a set of samples of the posterior distribution is obtained from which the Kullback-Leibler divergence can be calculated and, in turn, minimised. The algorithm has converged once the estimate for the mean and the estimate for the covariance are self-consistent.
The result of this algorithm is a set of samples of the approximated posterior distribution, implicitly containing the correlation structure between all model parameters. 
To apply this algorithm, we thus need 
\begin{enumerate}
    \item A model that allows drawing prior samples from a set of latent variables, taking into account the spatial correlation structure of the 3D gas distribution (the \textit{signal})
    \item A connection between the drawn gas realisation and the expected measurement data (the \textit{response})
\end{enumerate}

\subsection{Gas model and Signal}
Since the data from sky surveys has some fixed angular resolution, it is wise to represent the gas density on a grid that shares this property. If we chose to represent the gas density on a regular x-y-z grid (as in \cite{MertschCO, MertschHI}), voxels nearby would occupy almost half of the sky while voxels far away occupy an area on the sky much smaller than the available data resolution. In order to be consistent with the data resolution we thus choose to represent our signal data on a HEALPix-grid on angular direction and a logarithmic grid in radial direction. This ensures a high resolution nearby where we expect to be the most sensitive to the actual gas distribution. This choice will also make the response-function trivial as the otherwise costly line-of-sight integration reduces to a simple sum along the radial direction of the grid.

To model our prior, we generate samples of correlated lognormal random fields. These are obtained by drawing an -- initially white-noise -- sample $\xi(\Vec{x})$ of latent variables and correlating it using a method called Iterative Charted Refinement \cite{ICR} according to a Matérn-covariance function. We infer the parameters of this correlation structure at the same time as the gas density. The result is a correlated Gaussian random field $g(\Vec{x})$. Upon exponentiation, we obtain a lognormal correlation structure. This ensures positive gas densities while also allowing for large density differences as are expected to be present in the interstellar medium.
\begin{figure}
\centering
\begin{minipage}{.48\textwidth}
  \centering
  \includegraphics[width=.95\linewidth]{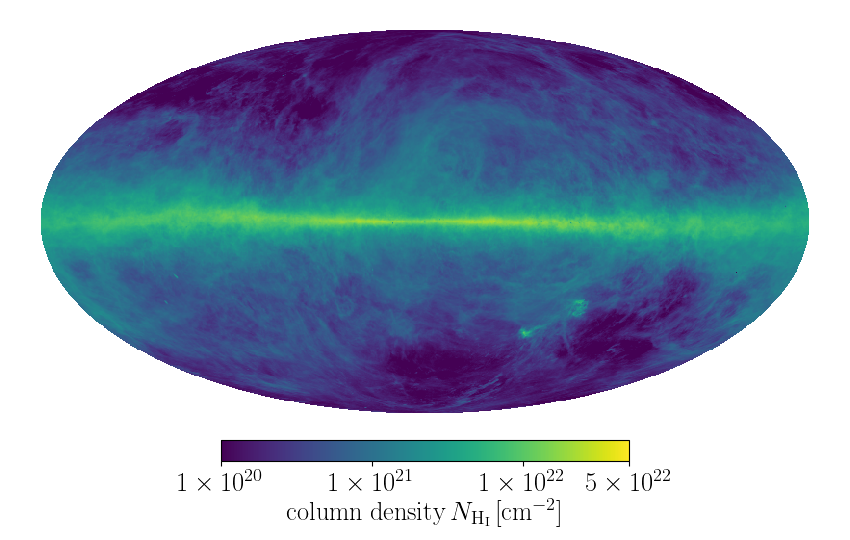}
  \captionof{figure}{HI column density map from the 21cm-data by the HI4PI collaboration \cite{HI4PI}}
  \label{fig:HIdata}
\end{minipage}%
\hfill
\begin{minipage}{.48\textwidth}
  \centering
  \includegraphics[width=.95\linewidth]{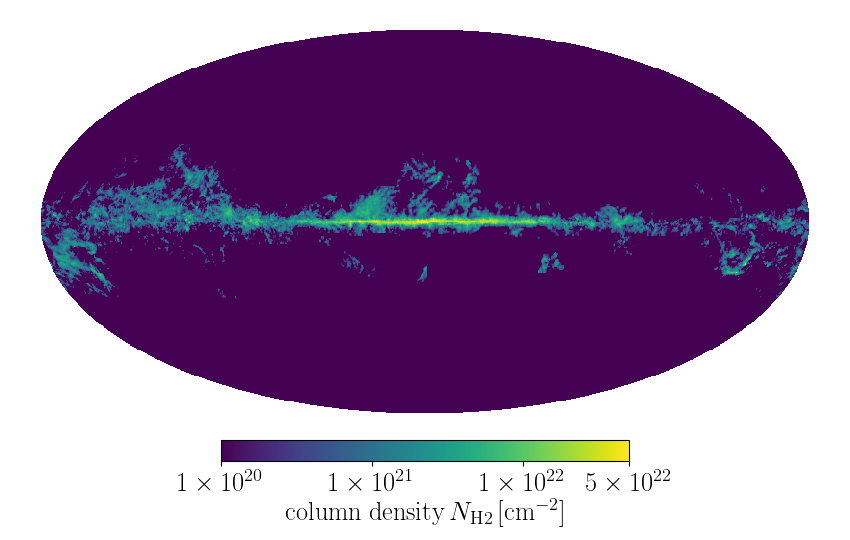}
  \captionof{figure}{H$_2$ column density map from the CO-emission-data compiled by Dame et al. \cite{DameCO}}
  \label{fig:COdata}
\end{minipage}
\end{figure}
This is not yet a very good prior assumption for gas in the galaxy as most of the gas is tightly constrained to the galactic disk which has a small scale height ($\approx$150\,pc) compared to its diameter ($\approx$15\,kpc). This can be immediately seen in the data-sets (figures \ref{fig:HIdata} and \ref{fig:COdata}): most of the gaseous emission is concentrated around latitude zero. The fidelity of the reconstruction can be increased by explicitly modelling the inhomogeneous large-scale variations.
We therefore multiply the correlated field with a profile in z-direction and in radial direction:
\begin{align}
    P_{z}(\Vec{x}) &= P_{z}(z) = \exp\left(\frac{-|z|}{z_h}\right),\\
    P_\text{rad}(\Vec{x}) &= P_\text{rad}(r_\text{gal}) = \exp\left(\frac{R_\text{cutoff}-r_\text{gal}}{R_\text{scale}}\right),\ \text{for}\ r_\text{gal}>R_\text{cutoff}, \ \text{else}\ 1\,.
\end{align}
Using this, we obtain
\begin{equation}
    \epsilon(\Vec{x}) = A \cdot P_{z}(\Vec{x})\cdot P_\text{rad}(\Vec{x})\cdot \exp\left(g(\Vec{x})\right)
\end{equation}
for HI and H$_2$ respectively.
For the HI-profile, we choose $z_h = z_h(r_\text{gal}) = 150\,\text{pc} \cdot \exp\left(\frac{r_\text{gal} - R_{\odot}}{9.8\,\text{kpc}}\right)$ for $r_\text{gal}>5\,\text{kpc}$, $R_\text{scale} = 3.15\,\text{kpc}$ and $R_\text{cutoff} = 7.0\,\text{kpc}$ as suggested by \cite{Kalberla}. For the H$_2$-profile, we use $z_h = 50\,\text{pc}$, $R_\text{scale} = 1.0\,\text{kpc}$ and $R_\text{cutoff} = 8.0\,\text{kpc}$. 
This does not prevent the inference from reconstructing fields that differ from this profile, but ensures that the drawn prior samples feature a galactic-disk-like gas distribution.

\subsection{Data and Response}

The second ingredient is the response function that connects the signal to the observation by modelling the generation of synthetic data from a signal sample. For simplicity, we work in the optically thin limit and ignore any absorption effects. In this case, the measured brightness temperature $T(\hat{n}, v)$ in some direction $\hat{n}$ Doppler-shifted by a velocity difference $v$ is related to the volume emissivity $\epsilon(\Vec{x})$ by a linear response map $R$ via
\begin{equation}
    T(\hat{n}, v) = R(\epsilon(\Vec{x})) = \int_0^\infty \text{d}r \epsilon(\Vec{x}) \delta(v - v_\text{LSR}(\Vec{x}))\,,
\end{equation}
where $v_\text{LSR}(\Vec{x})$ is the relative line-of-sight velocity at the position $(\Vec{x})$ in the local standard of rest as dictated by our velocity model and $r = |\Vec{x}|$ is the distance from Earth. We approximate the Dirac-delta by a Gaussian with a width of $\sigma_\text{HI} = 10\frac{\mathrm{km}}{s}$ \cite{Nakanishi2003} and $\sigma_\text{H$_2$} = 5\frac{\mathrm{km}}{s}$ \cite{Pohl_2008} in order to account for velocity dispersion inside gas clouds.

For the velocity model, we use a fixed component based on a smoothed particle hydrodynamics simulation by \cite{BEG03}, extended beyond 8\,kpc using a flat rotation curve. On top of this velocity field, we add another component computed as the gradient of a scalar velocity potential:
\begin{equation}
    v_\text{LSR}(\Vec{x}) = \Vec{v}_0(\Vec{x}) + \nabla S(\Vec{x})\,.
\end{equation}
This scalar field will be modelled as a correlated Gaussian random field and reconstructed at the same time as the gas emissivities . This opens the possibility for the model to adjust the velocity-field during the reconstruction. Additionally, we can hope to learn something about the true velocity field in areas, where the data is very constraining. However, there is no direct velocity-information in the data if one does not demand that the resulting gas densities should follow a certain correlation structure. Even then, this introduces many ambiguities as it greatly expands the possibilities, where gas clouds can be mapped. In the future, we will have to add additional data to either constrain the gas densities tighter (e.g. using correlations with dust \cite{DustEdenhofer}), thereby learning about the velocities or constrain the velocities tighter (e.g. using parallax information of masers or young stars \cite{Reid2019}), thereby learning about the gas densities.

\subsection{Noise and Likelihood}
Taking into account additive noise in the observations, we alter our model for the relation between brightness temperature and volume emissivity to 
\begin{equation}
    T(\hat{n}, v) = R(\epsilon(\Vec{x})) + n \,,
\end{equation}
where we assume the noise $n$ to be normal distributed and uncorrelated (white) with some diagonal covariance $N$. The likelihood can then be written as
\begin{equation}
    p(T|\epsilon) = \int\text{d}n\,p(T|\epsilon, n)p(n) = \int\text{d}n\, \delta(T-R(\epsilon)-n)\mathcal{G}(n, N) = \mathcal{G}(T-R(\epsilon), N)\,.
\end{equation}
\section{Results}
\begin{figure}
\begin{subfigure}{.32\textwidth}
  \includegraphics[scale=0.25]{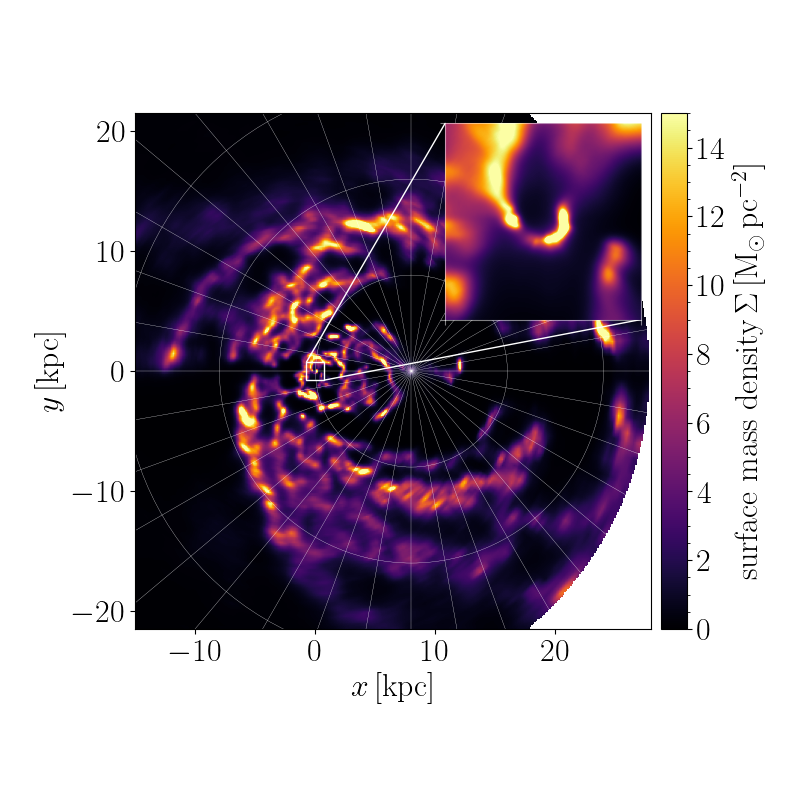}
    
\end{subfigure}%
\hfill
\begin{subfigure}{.32\textwidth}
  \includegraphics[scale=0.25]{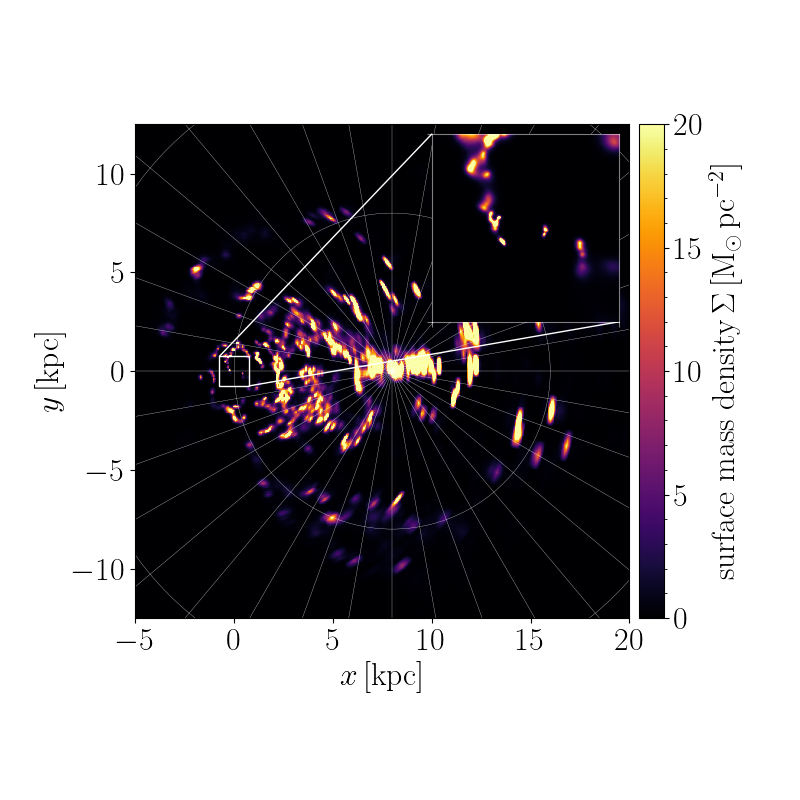}
    
\end{subfigure}%
\hfill
\begin{subfigure}{.32\textwidth}
    \includegraphics[scale=0.25]{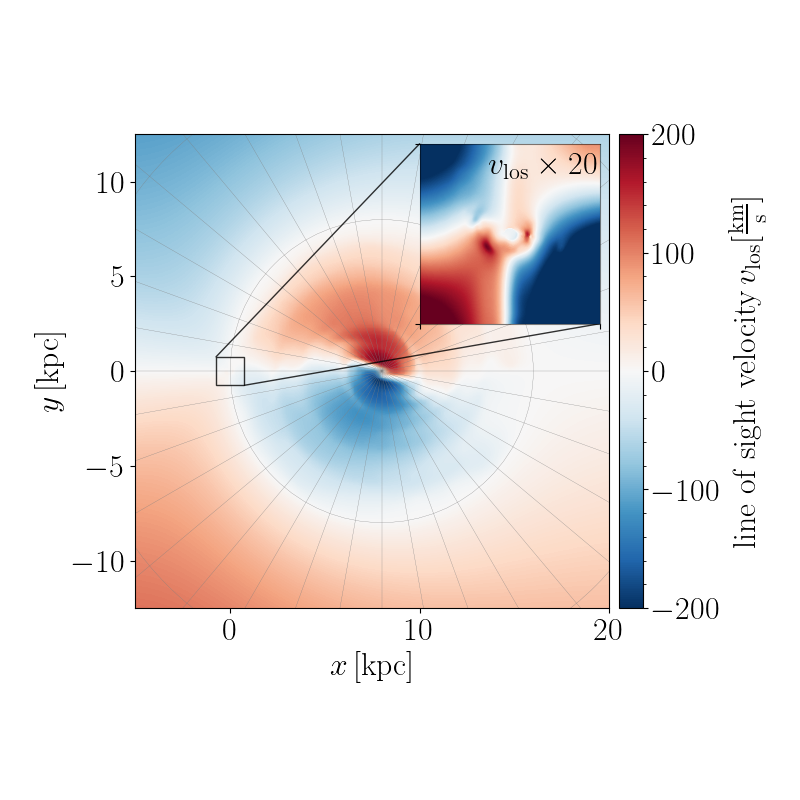}
    
\end{subfigure}%
\caption{Results of the inference with zoom-in on the local neighbourhood. Left panel: HI surface mass density. Middle panel: H$_2$ surface mass density. Right panel: Line-of-sight velocity at zero latitude}
\label{fig:HIzoomin}
\label{fig:H2zoomin}
\label{fig:vpotlsrzoomin}
\end{figure}

We run our reconstruction with a resolution of $\text{NSIDE}=32$ in angular direction and $500$ radial pixels between $r_\text{min}= 50 \,\text{pc}$ and $r_\text{max}= 28 \,\text{kpc}$. The sample-average surface mass densities resulting from the 3D-maps can be seen in figure \ref{fig:HIzoomin} for HI and H$_2$. This figure also shows the (partly reconstructed) sample-average line-of-sight velocity in a zero-latitude slice.
The reconstructed gas densities show disk-like structures of gas clusters with imprints of galactic arms that are particularly visible in the HI-gas reconstruction. Both gas reconstructions suffer from a set of problems that we will discuss in the following.
\subsection{Inferred HI gas density}
The outside of the solar circle is well populated with HI-gas, whereas in the inside of the solar circle, the distance ambiguity seems to be resolved very one-sidedly towards the nearer solution. This could be due to the logradial grid giving the algorithm the opportunity to place gas nearby with a much higher fidelity than far away. This can then reproduce the data with a much higher fidelity as well leading to a much higher likelihood. This could be tested and perhaps solved by e.g. modifying the grid to have a uniform resolution inside the solar circle or by increasing the total resolution until saturation.
The nearby gas shows a circular structure at the same radius as the reconstructed velocity as well as a tilted line-like structure in negative $x$-direction.
\subsection{Inferred \texorpdfstring{H$_2$}{H2} gas density}
The quality of reconstruction of H$_2$ appears to be worse than that of HI, mainly due to the fact that the gas is very concentrated at the plane $z=0$ and the amount of grid-points that are far away and very close to the galactic plane become very few. One can clearly see circular structures in the gas projection stemming from the (too) low angular resolution. Clearly visible is a bar-like structure in the galactic centre as well as two ''wall''-like structures in-between us and the galactic centre also seen in precious reconstructions on a regular grid \cite{MertschCO}. The excellent local resolution lets us see fine structures in the nearby gas showing a similar structure as the HI-gas in negative $x$-direction and small nearby clouds of gas towards the galactic centre.
%
\subsection{Inferred velocity field}
The reconstructed curl-free modification of the velocity prior is in general very small in amplitude and negligible for distances larger than $1\,\text{kpc}$. For distances smaller than that, there is an almost circular, positive (amplitude up to $10\frac{\mathrm{km}}{s}$) velocity correction being reconstructed. The position and amplitude coincide nicely with estimates for the expansion velocity of the local bubble \cite{GoodmanLocalBubble}. It is not clear, how much information about the velocity field itself is contained in the data but the combination of two data-sets having to respect the local correlation structure at the same time appears to provide at least some information.
\section{Conclusion}
We present new preliminary 3D-maps of galactic HI and H$_2$ inferred in conjunction using the same velocity field. We also present a partly reconstructed 3D line-of-sight velocity map featuring a circular structure with outwards-pointing velocities in the local neighbourhood. The ingredients for our inference were the HI4PI-survey from \cite{HI4PI} measuring 21cm-emission, the CO-survey compilation by \cite{DameCO} measuring rotational CO-transitions and a VI-algorithm capable of inferring millions of parameters \cite{MGVI}. We have assumed the optically thin limit. In the future we plan to improve upon these shortcoming by including additional data and thereby further constraining either the velocity field or the gas distributions, by lifting our assumption on the optical thinness of the gas; and by improving the angular resolution of our reconstructions.

\bibliographystyle{ieeetr}
\bibliography{proceedings}



\end{document}